\documentclass[aip,amsmath,amssymb,reprint,numerical]{revtex4-1}

\usepackage{amsfonts}
\usepackage{amsmath}
\usepackage{bm}
\usepackage{graphicx}
\usepackage{xcolor}
\usepackage{color}
\usepackage{amssymb}
\usepackage{natbib}

\def \beq {\begin{equation}}
\def \eeq {\end{equation}}
\def \bea {\begin{eqnarray}}
\def \eea {\end{eqnarray}}
\def \bfig {\begin{figure}}
\def \efig {\end{figure}}

\renewcommand{\thefootnote}{${}^\dag$}

\begin{document}

\title{Energy conversion in turbulent weakly-collisional plasmas: Eulerian Hybrid Vlasov-Maxwell simulations}

\author{O. Pezzi}
\email{oreste.pezzi@gssi.it}
\affiliation{Gran Sasso Science Institute, Viale F. Crispi 7, I-67100 L’Aquila, Italy}
\affiliation{INFN/Laboratori Nazionali del Gran Sasso, Via G. Acitelli 22, I-67100 Assergi (AQ), Italy}
\affiliation{Dipartimento di Fisica, Universit\`a della Calabria, I-87036 Cosenza, Italy}
\author{Y. Yang}
\affiliation{Southern University of Science and Technology, Shenzhen, Guangdong 518055, China}
\author{F. Valentini}
\affiliation{Dipartimento di Fisica, Universit\`a della Calabria, I-87036 Cosenza, Italy}
\author{S. Servidio}
\affiliation{Dipartimento di Fisica, Universit\`a della Calabria, I-87036 Cosenza, Italy}
\author{A. Chasapis}
\affiliation{Bartol Research Institute and Department of Physics and Astronomy, University of Delaware, Newark, DE 19716, USA}
\author{W.H. Matthaeus}
\affiliation{Bartol Research Institute and Department of Physics and Astronomy, University of Delaware, Newark, DE 19716, USA}
%
%
\author{P. Veltri}
\affiliation{Dipartimento di Fisica, Universit\`a della Calabria, I-87036 Cosenza, Italy}

\begin{abstract}
Kinetic simulations based on the Eulerian Hybrid Vlasov-Maxwell (HVM) formalism permit the examination of plasma turbulence with useful resolution of the proton velocity distribution function (VDF). The HVM model is employed here to study the balance of energy, focusing on channels of conversion that lead to proton kinetic effects, including growth of internal energy and temperature anisotropies. We show that this Eulerian simulation approach, which is almost noise-free, is able to provide an accurate energy balance for protons. The results demonstrate explicitly that the recovered temperature growth is directly related to the role of the pressure-strain interaction. Furthermore, analysis of local spatial correlations indicates that the pressure-strain interaction is qualitatively associated with strong-current, high-vorticity structures, although other local terms -such as the heat flux- weaken the correlation. These numerical capabilities based on the Eulerian approach will enable deeper study of transfer and conversion channels in weakly collisional Vlasov plasmas. \end{abstract}

\date{\today}

\maketitle

\section{Introduction}
\label{sect:intro}
The space physics community has become increasingly aware that plasma turbulence has significant impact on observed properties of solar wind and other accessible space plasmas \cite{Bruno&Carbone05, MatthaeusVelli11}. Given that the cascade rate towards smaller scales is measured\cite{MacBrideEA08, Sorriso-ValvoEA07,hadid2017energy} and  that an addition of proton internal energy is required in order to explain the observed temperature evolution within the inner heliosphere\cite{MattEA99-swh, HellingerEA13}, it is natural to study the {\it plasma dissipation} that must be present to reconcile these observations. Accordingly, kinetic dissipation of turbulence in weakly collisional plasmas has become a very active topic throughout heliospheric physics and several standard approaches have been employed. One approach is to assume that the notion of Ohmic dissipation carries over from the collisional case, and therefore to adopt as a proxy for dissipation the electromagnetic work on particles, namely ${\bm j} \cdot {\bm E}$ ($\bm j$ the electric current density, $\bm E$ the electric field) \citep{SundkvistEA07, WanEA15}. Another definition, especially adopted in simulation works, regards the role of inter-particle collisions, that introduce irreversibility into the system \citep{HowesEA06, TenBargeEA13-sheets, pezzi2016collisional, pezzi2017solarwind, pezzi2019protonproton}. Within this viewpoint, dissipation is associated with an entropy growth. Still different approaches study the dissipation related to peculiar mechanisms. Some examples are the dissipation of particular wave-modes, such as kinetic Alfv\'en\citep{salem2012identification,chen2019evidence} and whistler waves\citep{chang2015whistler,gary2016whistler}, and the heating related -directly or indirectly- to magnetic reconnection\citep{DrakeEA09,servidio2011statistical,servidio2012local,osman2012kinetic,osman2012intermittency,wu2013intermittent}. Magnetic reconnection in particular play a decisive role in energizing particles and dissipating energy in turbulent plasmas\citep{cerri2017reconnection}. 

In each of these approaches there is clearly an opportunity for physical insights, even if each approach may not capture all mechanisms that operate, or their mutual interactions. The concept of dissipation here advanced --and implemented in the analysis of HVM simulation of plasma turbulence-- differs in that it is based on the premise that pathways to dissipation may be identified based on separations of the effects of transport, scale transfer and energy conversion. In particular, one may uniquely characterize the conversion into internal energy, although this process is not related to an entropy growth.

Beginning with the HVM system, it is straightforward to derive energy balance relations \citep{krall1973principles}, which may be further analyzed to identify the basic channels of energy exchange
\cite{yang2017energyPOP, yang2017energyPRE} and the evolution of the pressure tensor\cite{DelSartoEA16, DelsartoPegoraro18}. One basic finding, reviewed in more detail below, is that the energy exchanged between the electromagnetic field and the proton bulk motions is due to the interaction of the electric field with the proton charge flux (current). Meanwhile, the flow kinetic energy may be exchanged with the internal energy through the combined action of the pressure-dilatation and the pressure-strain interactions. 

In the Vlasov energy balance, one also finds several energy transport terms that --although being quite important in dynamics-- do not contribute, on average, to converting energy from one form to another. In particular, transport terms cannot produce internal energy. These developments are based on elementary properties of the Vlasov equation, but their significance for collisionless turbulence has not always been fully recognized. 
Based on this taxonomy, one may proceed to analyze the contributions to these channels of energy conversion and transfer employing a number of approaches: by global averaging, by examining distribution functions, by computing joint distributions and correlations, and by studying real space maps to identify 
coherent structures \cite{yang2017energyPOP,yang2017energyPRE}.
Some of these analyses have also been implemented in MMS observations in the terrestrial magnetosheath \cite{chasapis2018energy}. This program of analysis of energy conversion channels is undertaken here by using HVM turbulence simulation. It is notable that, for HVM simulation, it is possible to reconcile a total energy balance with good accuracy, an undertaking not easily achieved using methods such as kinetic PIC. 
One may also implement scale averaging\citep{Germano92} to examine locality of energy transfer and the relative contribution of conversion processes at different scales \cite{yang2018scale, camporeale2018coherent}. This will be undertaken for HVM in a subsequent study. 

The paper is organized as follows. Section \ref{sect:model} introduces the theoretical model and the numerical setup of the numerical simulation performed here to implement the pressure-strain interaction {\it Pi-D} analysis, already adopted in recent studies \cite{yang2017energyPOP, yang2017energyPRE, yang2018scale, chasapis2018energy}. Then, in Sect. \ref{sect:overview}, we provide a general overview of the simulation results, in terms of production of a turbulent scenario. In Sect. \ref{sect:pid}, we describe several outcomes concerning the application of the {\it Pi-D} analysis on our simulation. Finally, in Sect. \ref{sect:concl}, we summarize the presented results.

\section{Theoretical model and channels of energy conversion}
\label{sect:model}

Within the HVM model \citep{valentini07,servidio2012local,perrone2013vlasov,servidio2015kinetic,valentini2016differential,cerri2017kinetic,perrone2018fluid}, protons are kinetic and electrons are a massless, isothermal fluid. Maxwell's equations are further approximated by neglecting the displacement current and by assuming charge neutrality. The Faraday's law and the generalized Ohm's law are then coupled to the proton Vlasov equation. Dimensionless HVM equations are:
\begin{eqnarray}
 &&\frac{\partial f_p}{\partial t} + {\bm v} \cdot \frac{\partial f_p}{\partial {\bm x} } +  \left ( {\bm E} + {\bm v} \times {\bm B} \right) \cdot \frac{\partial f_p}{\partial {\bm v}}=0, 
 \label{eq:HVMvlas} \\
&&\frac{\partial {\bm B}}{\partial t}\!=\! 
-\nabla\!\times\!{\bm E}  \label{eq:HVMfar} \\
 &&{\bm E}\!=- {\bm u}_p\times{\bm B}+ \frac{{\bm j}\times{\bm B}}{n_p} - \frac{\nabla P_e}{n_e} + \eta {\bm j}  \label{eq:HVMohm}
\end{eqnarray}
where $f_p({\bm x}, {\bm v}, t)$ is the proton distribution function (DF); ${\bm E}({\bm x}, t)$ and ${\bm B}({\bm x}, t)$ are respectively the electric and magnetic fields; ${\bm j}=\nabla\times{\bm B}$ is the current density; and $n_p=\int d^3v f_p$ and ${\bm u}_p = \int d^3v {\bm v} f_p/n_p$ are respectively the proton number density and bulk speed. The electron pressure $P_e=n_e T_e$ ($k_{_B}=1$) depends on spatial coordinates through the electron density $n_e=n_p$, while the electron temperature is only a parameter $T_e=T_{p,0}$, since electrons are isothermal. Equations (\ref{eq:HVMvlas}--\ref{eq:HVMohm}) have been normalized as follows. Time, velocities and lengths are respectively scaled to the inverse proton cyclotron frequency $\Omega_{cp}^{-1} = m_p c/eB_0$, to the  Alfv\'en speed $c_A = B_0 / \sqrt{4\pi n_{p,0} m_p}$, and to the proton skin depth $d_p = c_A/ \Omega_{cp}$; being $m_p$, $e$, $c$, $B_0$ and $n_{p,0}$ respectively the proton mass and charge, the light speed, the background magnetic field and the equilibrium proton density. 

The method described in the work of  \citet{yang2017energyPOP, yang2017energyPRE}, briefly revisited in the following, is based on the energy conservation equations, which --for the HVM model-- read as:
\begin{small}
\begin{eqnarray}
&&\frac{\partial \mathcal{E}^f_p}{\partial t} + \nabla \cdot \left({\bm u}_p \mathcal{E}^f_p + {\bm u}_p\cdot {\bm P}_p \right) = \left({\bm P}_p \cdot \nabla \right)\cdot{\bm u}_p + n_p{\bm u}_p \cdot {\bm E} \label{eq:enflow} \\
&&\frac{\partial \mathcal{E}^{th}_p}{\partial t} + \nabla \cdot \left({\bm u}_p \mathcal{E}^{th}_p + {\bm q}_p \right) = -\left({\bm P}_p \cdot \nabla \right)\cdot {\bm u}_p \label{eq:enth} \\ 
&& \frac{\partial \mathcal{E}^{m}}{\partial t} + \nabla \cdot \left({\bm E} \times {\bm B}  \right) =  - {\bm j} \cdot {\bm E} \label{eq:enmagn} 
\end{eqnarray}
\end{small}
where $\mathcal{E}^f_p= n_p {\bm u}_p^2/2$, $\mathcal{E}^{th}_p= \frac{1}{2} \int d^3v f_p \left({\bm v}-{\bm u}_p\right)^2 = 3n_pT_p/2$ and $\mathcal{E}^{m}= {\bm B}^2/2$ are respectively the flow, internal, 
and magnetic energy densities. Note that the proton total energy is $\mathcal{E}_p = \frac{1}{2}\int d^3v f_p {\bm v}^2 = \mathcal{E}^f_p + \mathcal{E}^{th}_p$. Moreover, the kinetic temperature $T_p$, implicitly 
defined as usual through the VDF $2^{nd}$-order moment, does not imply thermal equilibrium.

In Eqs. (\ref{eq:enflow}--\ref{eq:enmagn}), the pressure tensor ${\bm P}_p$ and the heat flux ${\bm q}_p$ are defined as follows: 
\begin{eqnarray}
 && {\bm P}_p = \frac{1}{2} \int d^3v ({\bm v}-{\bm u}_p) ({\bm v}-{\bm u}_p) f_p \label{eq:Presstens} \\ 
 && {\bm q}_p = \frac{1}{2} \int d^3v ({\bm v}-{\bm u}_p)^2 ({\bm v}-{\bm u}_p) f_p \label{eq:heatflux}
\end{eqnarray}

\begin{figure*}[!htb]   
\includegraphics[width=\textwidth]{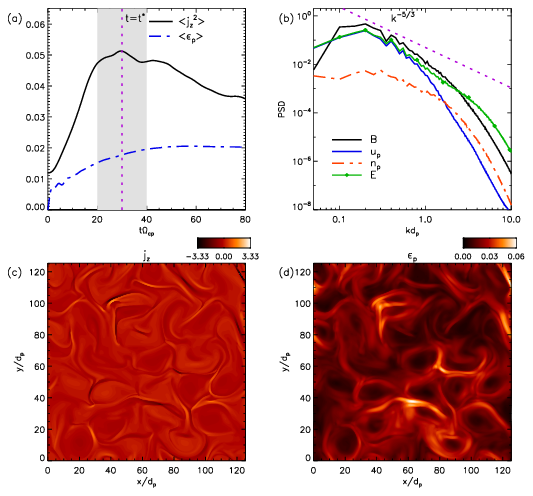}
\caption{(a) Temporal evolution of $\langle j_z^2\rangle(t)$ (black solid); $\langle\epsilon_p\rangle$(t) (blue dash-dotted). The purple dashed line indicates the time at which turbulence activity is strongest $t=t^*=30\Omega_{cp}^{-1}$, being $t^* = 0.6 \tau_{nl}$. (b) Omni-directional perpendicular PSDs of the magnetic field (black); bulk speed (blue); number density (red dash-dotted) and of the electric field (green dotted). The purple dashed line reports the Kolmogorov expectation $k^{-5/3}$ as reference. PSDs are computed by averaging on the temporal range $t\Omega_{cp}\in[20,40]$, that corresponds to a nearly constant $\langle j_z^2\rangle$ and is shown in panel (a) with a gray shaded area. (c--d) Contour plots of $j_z(x,y)$ (c) and $\epsilon_p(x,y)$ (d) at the temporal instant $t=t^*$.} 
\label{fig:simOV}
\end{figure*}

To appreciate the conversions between different type of energy, Eqs. (\ref{eq:enflow}--\ref{eq:enmagn}) are averaged on the spatial domain to get:
\begin{eqnarray}
 &&\frac{\partial \langle\mathcal{E}^f_p\rangle}{\partial t} = \langle {\bm j}_p \cdot {\bm E}\rangle + \langle \left({\bm P}_p \cdot 
\nabla \right)\cdot {\bm u}_p\rangle \label{eq:enflowav} \\
&&\frac{\partial \langle\mathcal{E}^{th}_p\rangle}{\partial t} =-\langle \left({\bm P}_p \cdot \nabla \right)\cdot {\bm 
u}_p\rangle \label{eq:enthav}\\ 
&& \frac{\partial \langle\mathcal{E}^{m}\rangle}{\partial t} =-\langle {\bm j} \cdot {\bm E} \rangle \label{eq:enmagnav} 
\end{eqnarray}
where ${\bm j}_p = n_p{\bm u}_p$ ($e=1$ in scaled units) and all divergence terms vanish due to the spatial boundary conditions. Note that the sum of the terms at right-hand sides of Eqs. (\ref{eq:enflowav}--\ref{eq:enmagnav}) is not zero, but reduces to the electromagnetic work performed on electrons ($\sim \langle {\bm j}_e\cdot {\bm E} \rangle$). 
This term, for the HVM algorithm, is 
determined by the hybrid plasma approximation
described above, and is not reducible 
to a simple time derivative. Eqs. (\ref{eq:enflowav}--\ref{eq:enmagnav}) suggest that the term $\langle {\bm j}_p \cdot {\bm E} \rangle$ controls the transfer between flow and magnetic energy, while $\langle \left({\bm P}_p \cdot \nabla \right)\cdot {\bm u}_p\rangle$ transforms flow energy into internal energy \citep{yang2018scale}. 
This last term can be further decomposed as:
\begin{equation}\label{eq:pid}
 \left({\bm P}_p \cdot \nabla \right)\cdot {\bm u}_p = P_p \theta_p + {\bm \Pi}_p : {\bm D}_{p}
\end{equation}
where $P_{p,ij} = P_p \delta_{ij} + \Pi_{p,ij}$, $P_p=P_{p,ii}/3$, $\theta_p=\nabla \cdot {\bm u}_p$ and $D_{p,ij}=\left(\partial_j u_{p,i} + \partial_i u_{p,j} \right)/2 - \theta_p \delta_{ij}/3$. In the last expression,  $\delta_{ij}$ is the Kronecker's delta and $\partial_i$ denotes the partial derivatives with respect to the $i$--th spatial coordinate. The two terms at the right-hand side of Eq. (\ref{eq:pid}) can be respectively associated with plasma dilatations and compressions and with the trace-less pressure tensor (including also off-diagonal pressure terms). Hereafter, $P_p\theta_p$ and ${\bm \Pi}_p : {\bm D}_{p}$ are briefly denoted as the {\it P-$\theta$} and {\it Pi-D} terms, respectively. Both terms appear with a negative sign in Eq. (\ref{eq:enthav}), thus implying that negative values of $\langle P_p\theta_p \rangle$ and $\langle {\bm \Pi}_p : {\bm D}_{p} \rangle$ have the effect of increasing the proton internal energy. Moreover, both terms are not positive definite, hence their signs are decisive to discriminate the direction of the energy transfer between flow and internal energy.

The idea adopted in recent studies\cite{yang2017energyPOP, yang2017energyPRE, yang2018scale} is that the {\it Pi-D} term can provide significant insights on the mechanisms that transfer energy towards smaller scales, where energy is eventually dissipated. Indeed, in a collisional plasma (MHD framework), the traceless pressure-tensor terms are usually related to viscous dissipation \citep{braginskii1965transport}. Investigating whether this connection holds also in a weakly-collisional system can be hence useful to understand the nature of dissipation, although --within the HVM model-- collisions are not modeled and the system is formally reversible\citep{pezzi2016collisional, pezzi2017solarwind, pezzi2019protonproton}. 

\section{Overview of the Numerical Simulations}
\label{sect:overview}

Equations~(\ref{eq:HVMvlas}--\ref{eq:HVMohm}) are integrated in a 2.5D--3V phase space domain. We retain all velocity space directions, while physical-space vectors are three-dimensional but depend only on the two spatial coordinates $(x,y)$. This double-periodic spatial domain, whose size is $L_x=L_y=L=2\pi\times 20 d_p$, is discretized with $N_x=N_y=512$ grid-points. The velocity domain is discretized with $N_{v_x}=N_{v_y}=N_{v_z}=71$ points in the range $v_{j}=\left[-5 v_{th,p},5v_{th,p}\right]\;(j=x,y,z)$. Boundary conditions impose $f(v_j>5 v_{th,p})=0$, where the proton thermal speed  $v_{th,p}=\sqrt{k_{_B} T_{p,0}/m }$ is directly related to the Alfv\'en speed through $\beta_p=2 v_{th,p}^2/c_A^2= 2$. 

The proton VDF is initially Maxwellian, with uniform unit density. A uniform background out-of-plane magnetic field ${\bm B_0} = B_0 {\bm e}_z$ ($B_0=1$) is initially imposed. The equilibrium is perturbed by imposing an initial $2D$ spectrum of Fourier modes, equi-partitioned between proton bulk velocity and magnetic field, and having an amplitude such that $\delta B_{rms}/B_0 = \delta u_{rms}/c_A= 1/2$. The energy is injected with random phases and in the wave-number range $k\in[0.1,0.3]$, being $k=m k_0$ with $ 2\leq m \leq 6$ and $k_0=2\pi/L$. Neither density fluctuations nor parallel perturbations are introduced ($\delta n_p=\delta u_{p,z} = \delta B_z = 0$). A small resistivity $\eta=1.5\times 10^{-2}$ is accurately introduced to avoid numerical instabilities.

Before focusing on the {\it Pi-D} analysis, we conclude this section by describing the simulation results concerning the onset of turbulence at proton inertial scales. 

Owing to the presence of strong nonlinearities, the initial relatively large-scale perturbations couple and produce fluctuations at smaller scales. This cascade-like activity can easily be appreciated by examining the evolution of $\langle j_z^2\rangle$ [Fig. \ref{fig:simOV}(a)], that represents a standard proxy for highlighting the presence of nonlinear couplings. The $\langle j_z^2\rangle$ peak, occurring at  $t^*=30\Omega_{cp}^{-1}$, is associated with the most intense turbulent activity in the numerical simulation \citep{servidio2012local}. Note that the time $t^*$ corresponds to about $0.6\tau_{nl}$, where the nonlinear time is $\tau_{nl}=l_c/\delta u(l_c)$, $l_c$ is the turbulence correlation length and $\delta u(l_c)$ is the fluctuations amplitude at $l_c$. In our case, $l_c\simeq L/5$ and $\delta u(l_c)\simeq \delta u_{rms}$. The generation of small-scale fluctuations can be recognized by looking at the omni-directional (perpendicular) power spectral densities (PSDs), reported in Fig. \ref{fig:simOV}(b), as a function of $k d_p$. PSDs have been computed by averaging on the temporal range $t\Omega_{cp}\in[20,40]$ (shaded area in Fig. \ref{fig:simOV}(a)), where $\langle j_z^2\rangle$ is nearly constant. This time interval corresponds to a quasi-stationary regime of turbulence. For about a decade of scales larger than the proton inertial scales, an inertial range is observed, wherein magnetic fluctuations are dominant and show a power-law-like PSD, which is reminiscent of the $-5/3$ Kolmogorov slope. At smaller scales, the electric field power increases as well as compressibility \citep{Bale05}.

The real-space picture of out-of-plane electric current $j_z(x,y)$ at $t=t^*$, is reported in Fig. \ref{fig:simOV}(c). This contour plot confirms an highly structured and inhomogeneous current density pattern, with strong current sheets and magnetic islands. In, or near, these strong current sheets, which themselves are embedded in a dynamic and strongly turbulent region, complex phenomena such as magnetic reconnection, energy dissipation and other kinetic effects \citep{servidio2015kinetic, retino2007insitu, osman2011evidence, karimabadi2013coherent} are found. Intense current regions are often in close proximity to regions in which the plasma VDF is far from the equilibrium Maxwellian shape \citep{drake2003formation,drake2010magnetic,greco2012inhomogeneous,swisdak2016quantifying,pezzi2017colliding}. The presence of non-Maxwellian features in the proton VDF can be also associated with the enstrophy velocity-space cascade \citep{servidio2017magnetospheric, cerri2018dual, pezzi2018velocityspace}. To quantify deviations of the proton VDF from the associated Maxwellian $f_M$, we here adopt the $\epsilon_p$ parameter \cite{greco2012inhomogeneous}:
\begin{equation}
 \epsilon_p({\bm x},t) = \frac{1}{n_p}\sqrt{\int{(f_p-f_{_M})^2 d^3v}}.
 \label{eq:epsp}
\end{equation}
where $f_{_M}$ has the same density, bulk speed and temperature of $f_p$. The temporal evolution of $\langle \epsilon_p\rangle(t)$, reported in Fig. \ref{fig:simOV}(a), highlights that the generation of non-Maxwellian features grows together with the increase of the current activity (i.e. magnetic field gradients and small-scale fluctuations). The $\epsilon_p$ parameter finally saturates at a nearly constant value. The contour plot of $\epsilon_p(x,y)$ at $t=t^*$, reported in Fig. \ref{fig:simOV}(d), demonstrates a qualitative correlation of $\epsilon_p(x,y)$ and $j_z(x,y)$.

\begin{figure*}[!htb]
\includegraphics[width=0.75\columnwidth]{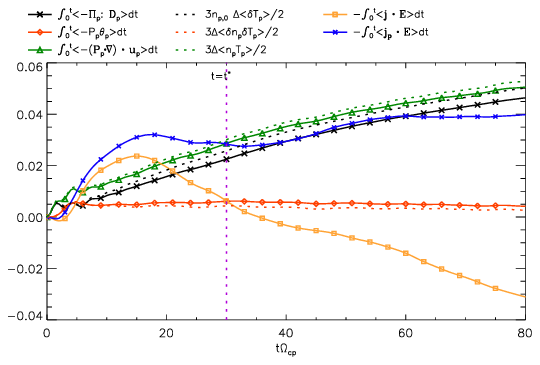}
\caption{Temporal evolution of the cumulative integral of: $\langle-{\bm \Pi}_p : {\bm D}_{p}\rangle$ (black solid with crosses); $\langle -P_p\theta_p\rangle$ (red solid with rhombuses); and $\langle -\left( {\bm P}_p \cdot \nabla \right)\cdot {\bm u}_p\rangle$ (green solid with triangles), that is the sum of the first two terms. Black, red and green dot-dashed lines show $3 n_{p,0} \Delta \langle \delta T_p \rangle/2 $; $3\Delta \langle \delta n_p \delta T_p \rangle/2$; and $3 \Delta \langle n_p T_p\rangle/2$, respectively. The orange and blue solid lines indicates the cumulative integral of $\langle -{\bm j}\cdot {\bm E}\rangle$ and of $\langle -{\bm j}_p\cdot {\bm E}\rangle$, respectively.} 
\label{fig:timeev}
\end{figure*}

\section{The {\it Pi-D} analysis: energy conversion and temperature increase} \label{sect:pid}

\begin{figure*}[!hbt]
\includegraphics[width=15cm]{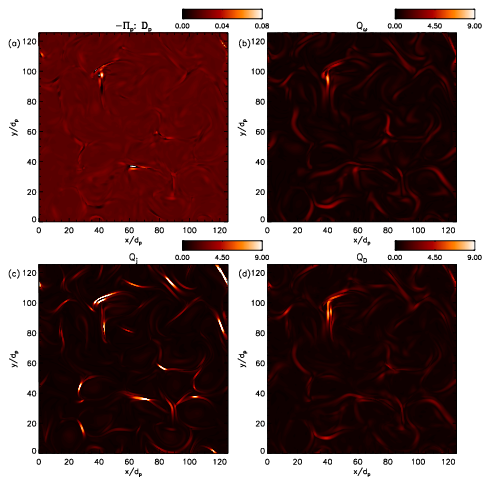}
\caption{Contour plots of (a) $-{\bm \Pi}_p : {\bm D}_p$;  
(b) $Q_{\bm \omega}= {\bm \omega}^2/4 \langle {\bm \omega}^2\rangle $; (c)   $Q_{\bm j}= {\bm j}^2/4 \langle {\bm j}^2\rangle   $; and (d) $Q_D =  {\bm D}_{p}: {\bm D}_{p}/ 4\langle {\bm D}_{p}: {\bm D}_{p}\rangle   $ (d).} 
\label{fig:maps}
\end{figure*}

\subsection{Bookkeeping the internal energy}
In order to understand the energy transfer mechanisms occurring in a turbulent plasma, it is useful to apply the {\it Pi-D} method on the simulation described in Section  \ref{sect:model}.

By integrating on time Eq. (\ref{eq:enthav}) and using Eq. (\ref{eq:pid}), one gets:
\begin{equation}
 \Delta \langle\mathcal{E}^{th}_p\rangle (t) = - \int_0^t d\tau \langle {\bm \Pi}_p : {\bm D}_{p} \rangle(\tau) - \int_0^t d\tau \langle P_p\theta_p\rangle(\tau) 
\label{eq:enthav_CI}
\end{equation}
where $\Delta$ denotes the variation with respect to the initial values. Figure \ref{fig:timeev} reports the temporal evolution of the cumulative integrals of $\langle- {\bm \Pi}_p : {\bm D}_{p} \rangle$; $-\langle P_p\theta_p\rangle$ ; and $\langle -\left( {\bm P}_p \cdot \nabla \right)\cdot {\bm u}_p\rangle$. These integrals have been numerically evaluated through the trapezoidal rule. The {\it Pi-D} term dominates over the pressure-dilatation term in the internal energy increase,
suggesting that the cascade process remains quasi-incompressible. Since Eulerian simulations are almost noise-free, we can also report the evolution of the internal energy $\Delta  \langle\mathcal{E}^{th}_p\rangle (t) = 3 \Delta \langle n_p T_p\rangle/2$. According to Eq. (\ref{eq:enthav}), the internal energy variation should be exactly equal to the cumulative integral of $\langle -\left( {\bm P}_p \cdot \nabla \right)\cdot {\bm u}_p\rangle$. The slight difference between these two quantities is mainly due to the level of accuracy for the energy conservation recovered in the numerical simulation ($\sim0.5\%$).

\begin{figure*}[!htb]
\includegraphics[width=12cm]{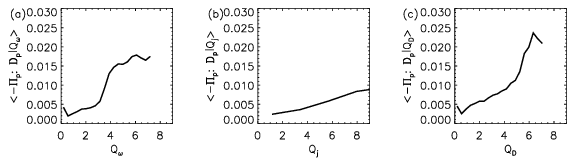} \caption{Conditional averages of $-{\bm \Pi}_p : {\bm D}_p$ term with respect to $Q_{\bm \omega}$ (a), $Q_{\bm j}$ (b) and $Q_{D}$ (c). } 
\label{fig:condav}
\end{figure*}

It can be also shown that the {\it Pi-D} term has a direct effect on the temperature growth. This effect can be highlighted by writing $n_p=n_{p,0} + \delta n_p$ and $T_p=T_{p,0} + \delta T_p$, where $\langle \delta n_p\rangle =0$ since the total mass is well preserved. Therefore:
\begin{equation}
    \Delta \langle\mathcal{E}^{th}_p\rangle (t) = \frac{3}{2} \left( n_{p,0} \Delta \langle \delta T_p \rangle + \Delta \langle \delta n_p \delta T_p \rangle \right)
\label{eq:enth2}
\end{equation} 
The first term in this expression, namely $ 3 n_{p,0} \Delta \langle \delta T_p \rangle/2$,  is clearly connected to temperature growth that uncorrelated from density fluctuations. It can be almost superposed 
on the {\it Pi-D} term evolution, the slight discrepancy arguably due to the simulation accuracy. This behavior can be easily explained by deriving, from Eq. (\ref{eq:enth}), the evolution equation for the temperature $T_p$:
\begin{equation}
\frac{3}{2} \left[ \frac{\partial T_p}{\partial t} + \nabla \cdot \left({\bm u}_p T_p \right)\right] = - \frac{1}{n_p} \left( P_p\theta_p + {\bm \Pi}_p : {\bm D}_p + \nabla \cdot {\bm q}_p \right) 
\label{eq:tempev}
\end{equation}
By averaging on the spatial domain and by considering a weakly-compressible case as the one analyzed here (density fluctuations always remain quite weak being, at $t=t^*$, $\delta n_{p,rms}/n_{p,0} \sim 7.7 \times 10^{-2}$), it can be easily deduced that the dominant term in Eq. (\ref{eq:tempev}) is the {\it Pi-D} one. Hence, one gets:
\begin{equation}
\frac{3}{2} n_{p,0} \frac{\partial \langle T_p \rangle}{\partial t}  \simeq - \langle {\bm \Pi}_p : {\bm D}_p \rangle  \; \; ,
\label{eq:tempev2}
\end{equation}
that fully clarifies the behavior observed in Fig. \ref{fig:timeev}.

On the other hand, the evolution of the second term in Eq. (\ref{eq:enth2}), namely $3\Delta \langle \delta n_p \delta T_p \rangle/2$, is similar to {\it P-$\theta$}, as it can be expected since eventual correlations between density and temperature can be understood in terms of compression and dilatation.

Note also that, in the current simulation, the average temperature increases without showing a saturation regime. The evolution of correlated density-temperature fluctuations instead saturates in the first stages of the numerical simulation. This is probably due to the finite level of compressive-like structures contained into the initial perturbations. This effect may possibly be understood in terms if nearly-incompressible MHD theory \citep{Zank92}. Detailed analysis of the effects of different initial conditions, namely by i) introducing a different level of initial compressibility and by ii) changing the value of $\beta_p$, that also regulates the level of compressive activity \citep{bale2009magnetic}, will be the content of a future work.

\subsection{Role of ${\bm j} \cdot {\bm E}$}

In the initial stage of the simulation and up to the peak of the turbulence ($t\lesssim t^*$) 
$\int_0^t dt\langle - {\bm j}\cdot {\bm E} \rangle$ is positive, 
associated with the initial 
increase of magnetic fluctuation energy at the expense of flow energy (a feature familiar from MHD simulations).  This evolution does not compare well with the temperature growth in Fig. \ref{fig:timeev}, 
suggesting that, despite being adopted frequently as a proxy for heating and dissipation, the electromagnetic work is not globally correlated with the temperature increase. This information is already partially contained in Eqs. (\ref{eq:enflowav}--\ref{eq:enmagnav}), even though --in principle-- there could be transfers between magnetic and internal energy, through the intermediary of the kinetic energy. 

For the sake of completeness, 
in Fig. \ref{fig:timeev}, we also report the cumulative integral of $\langle - {\bm j}_p\cdot {\bm E} \rangle$. Compared 
to the integral of total 
$\langle - {\bm j}\cdot {\bm E} \rangle$,
the difference represents 
the physical work that is done on electrons. 
It is worth highlighting that here we are focusing on global (averaged) quantities, while ${\bm j}\cdot {\bm E}$ is locally important to identify sites 
of kinetic activity and heating occurs \cite{Zenitani:etal:2011,Wan:etal:2012,GrecoEA12-kinetic}. 

\subsection{Spatial Distributions and Correlations}
Figure \ref{fig:maps} reports the contour plots of $-{\bm \Pi}_p : {\bm D}_p$ (a), $Q_{\bm \omega} = {\bm \omega}^2/4 \langle {\bm \omega}^2\rangle $ (b), $Q_{\bm j} = {\bm j}^2/4 \langle {\bm j}^2\rangle $ (c) and $Q_D =  {\bm D}_{p}: {\bm D}_{p}/ 4\langle {\bm D}_{p}: {\bm D}_{p}\rangle$ (d), at the time instant $t=t^*$. $Q_{\bm \omega}$, $Q_{\bm j}$, and $Q_D$ estimate the relative strength of rotation, current sheets, and strain, respectively. 

Our finding supports the recent works \cite{yang2017energyPOP, yang2017energyPRE, yang2018scale} where a coarse-grained 
correlation between intense {\it Pi-D} regions as well as strong current structures and and high vorticity and velocity strain regions has been described. These correlations can be also appreciated by looking at Fig. \ref{fig:condav}, where the conditional averages of $-{\bm \Pi}_p : {\bm D}_p$ with respect to $Q_{\bm \omega}$ (a), $Q_{\bm j}$ (b) and $Q_{D}$ (c) are reported: stronger {\it Pi-D} event regions are found near regions of strong fluid vorticity and strain, with a weaker correlation evident near regions of current enhancement. 
Although we are not modeling irreversible dissipation, the correlation between {\it Pi-D} and $Q_{D}$ --usually related to viscosity in collisional plasmas-- 
suggests a connection between 
two intrinsically different physical systems (collisionless and collisional).

\begin{figure}[!htb]
\includegraphics[width=7.5cm]{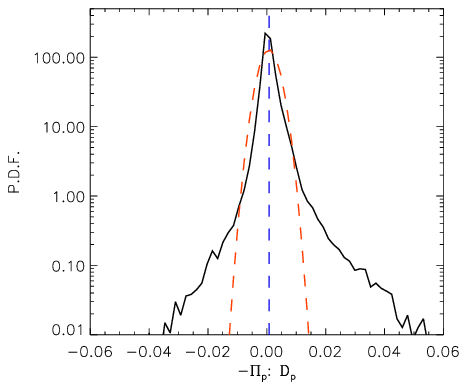}
\caption{Probability Density Function (P.D.F.) of $-{\bm \Pi}_p : {\bm D}_p$ (black) and Gaussian expectation (red dashed), at $t=t^*$. The blue dashed vertical line is the average value of the P.D.F.: $\langle -{\bm \Pi}_p : {\bm D}_p \rangle = 7.5 \times 10^{-4}$. } 
\label{fig:PDF}
\end{figure}

To further support the idea that the plasma heating is an inhomogeneous process \citep{osman2011evidence}, Fig. \ref{fig:PDF} shows the probability density function (P.D.F.) of $-{\bm \Pi}_p : {\bm D}_p$. The {\it Pi-D} P.D.F. is far different from the Gaussian expectation, confirming the presence of intermittency. The {\it Pi-D} P.D.F. is also skewed, or slightly elongated in the positive direction of $-{\bm \Pi}_p : {\bm D}_p$, i.e. in the direction of net increase of the internal energy, as indicated by the positive sign of the $-{\bm \Pi}_p : {\bm D}_p$ average.

\begin{figure*}[!htb]
\begin{minipage}{0.3 \textwidth}
\includegraphics[scale=0.65]{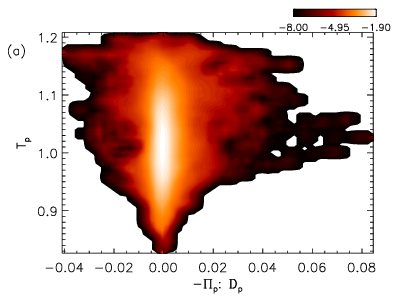}
\end{minipage}
\begin{minipage}{0.3 \textwidth}
\includegraphics[scale=0.65]{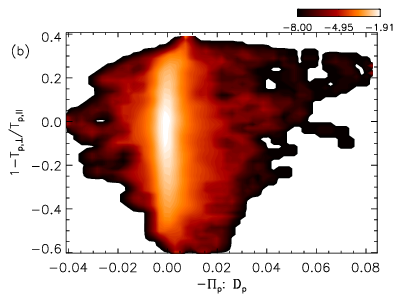}
\end{minipage}
\begin{minipage}{0.3 \textwidth}
\includegraphics[scale=0.65]{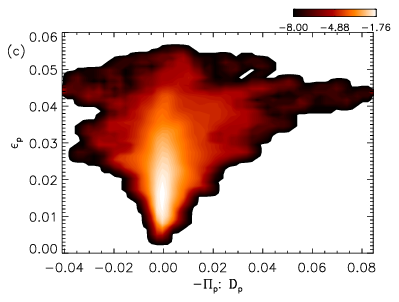}
\end{minipage}
\caption{Joint P.D.F. of $T_p$ and $-{\bm \Pi}_p : {\bm D}_p$ (a); $1-T_\perp/T_\parallel$ and $-{\bm \Pi}_p : {\bm D}_p$ (b); and $\epsilon_p$ and $-{\bm \Pi}_p : {\bm D}_p$ (c). Contour levels are in log scale.}
\label{fig:JPDF}
\end{figure*}

\begin{figure*}[!htb]
\includegraphics[width=15cm]{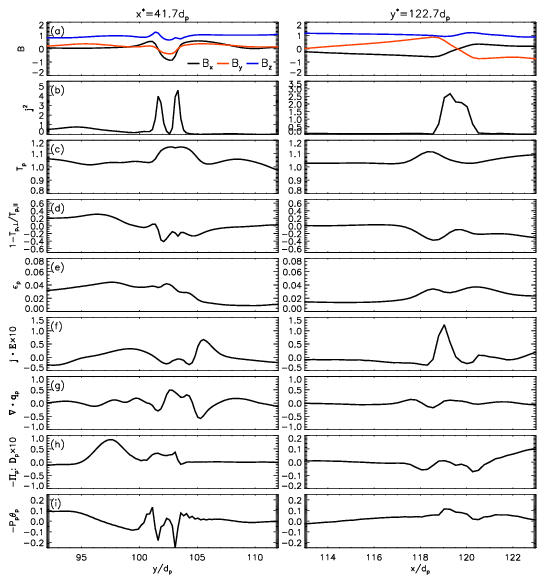}
\caption{One dimensional profile of
 several quantities, selected at $x=41.7 d_p$ (left column) and $y=122.7 d_p$ (right column). Each row 
reports, from top to bottom: (a) magnetic field components $B_x$ (black), $B_y$ (red) and $B_z$ (blue); (b) current density ${\bm j}^2$ ; (c) proton temperature $T_p$; (d) temperature anisotropy $1-T_{p,\perp}/T_{p,\parallel}$; (e) $\epsilon_p$; (f) ${\bm j}\cdot{\bm E}$; (g) $\nabla \cdot {\bm q}_p$; (h) $-{\bm \Pi}_p : {\bm D}_p$; and (i) $-P_p\theta_p$.} 
\label{fig:1Dcut}
\end{figure*}

Since Eulerian Vlasov simulations are almost noise-free, we can reveal whether the {\it Pi-D} term is correlated with kinetic characteristics of the proton VDF. This aspect can be also addressed with PIC simulations, but only when a quite large number of particles per cell are adopted \citep{camporeale2011dissipation,franci2015solarwind,franci2018solarwind}; otherwise several of the features shown below are masked due to 
the presence of numerical noise \citep{pezzi2017colliding}. Figure \ref{fig:JPDF} reports the joint P.D.F. of the total temperature $T_p$ (a), temperature anisotropy $1-T_{p,\perp}/T_{p,\parallel}$ (b) and non-Maxwellianity parameter $\epsilon_p$ (c) with $-{\bm \Pi}_p : {\bm D}_p$. Each joint P.D.F. reveals an asymmetry in the positive direction $-{\bm \Pi}_p : {\bm D}_p$. A subtle correlation between regions of strong {\it Pi-D} and areas where plasma is locally warmer and non-Maxwellian can be seen. However, the {\it Pi-D} term is broadly distributed 
and therefore 
not strongly correlated with proton 
temperature heating nor with the plasma non-Maxwellianity, in part due to 
transport effects. 
Indeed, strong $\epsilon_p$, $T_p$ and $1-T_{p,\perp}/T_{p,\parallel}$ are also associated with very weak $-{\bm \Pi}_p : {\bm D}_p$ regions. 

The {\it Pi-D} term reveals hence regions where kinetic processes may be at work, but the viceversa is not true. This is probably caused by the fact that other terms can be locally important and may break the correlation. With regard to this last comment on correlations between {\it Pi-D} and kinetic signatures, it should be borne in mind that, when examining the {\it Pi-D} term at the local level or using {\it in-situ}  measurements\citep{chasapis2018energy}, transport terms may also play a role. Some of these terms, that are related to divergence, move energy from one location to another without globally changing the form of energy.

Furthermore, both the magnetic field and the spatial derivatives of the bulk speed enter into the dynamics of the pressure tensor \cite{DelSartoEA16}. Consequently, various kinetic and fluid signatures of dissipation, such as {\it Pi-D}, current, ${\bm j}\cdot {\bm E}$, vorticity, temperature anisotropy and non-Maxwellianity, may be coarsely or regionally correlated, but pointwise offset from one another. This type of regional clustering of quantities related to dissipation has been observed in several simulation studies \cite{Wan:etal:2012,servidio2015kinetic,yang2018scale,camporeale2018coherent} and is an important facet of intermittency in plasma turbulence. 

\subsection{Role of local transport terms}
To further support that local effects are important in the energy transfer towards smaller scales, where dissipation is thought to occur, we conclude this section by selecting two vertical cuts in the spatial domain of the numerical simulation that cross strong current sheet. 

Figure \ref{fig:1Dcut} shows these one-dimensional cuts, taken at $x=41.7 d_p$ (left column) and $y=122.7 d_p$ (right column). Each row reports, from top to bottom, (a) magnetic field components $B_x$, $B_y$ and $B_z$; (b) current density ${\bm j}^2$; (c) proton temperature $T_p$; (d) temperature anisotropy $1-T_{p,\perp}/T_{p,\parallel}$; (e) non-Maxwellian measure $\epsilon_p$; (f) ${\bm j}\cdot{\bm E}$; (g) the heat flux divergence $\nabla \cdot {\bm q}_p$; (h) $-{\bm \Pi}_p : {\bm D}_p$; and (i) $-P_p\theta_p$.

In both events, at least one component of the magnetic field changes its sign and the current density has a significant growth, characterized by a double-peaked structure in the left column, as recently observed also in Earth's magnetosheath \citep{chasapis2018energy}. The proton temperature also increases close to the current density peaks. In the left-column panel, plasma is hotter inside the current sheet, while in the right-column panel, plasma heating is slightly displaced from the current sheet \cite{servidio2015kinetic}. 

Both proxies of non-Maxwellianity reported in Fig. \ref{fig:1Dcut}, namely $1-T_{p,\perp}/T_{p,\parallel}$ (d) and $\epsilon_p$ (e), present a broad structure where both parameters are significantly non-null. The presence of such large-scale structures of non-Maxwellianity also causes the lack of correlation between the non-Maxwellianity parameters and {\it Pi-D}. Super-imposed to these large-scale structures, quite strong peaks of $1-T_{p,\perp}/T_{p,\parallel}$, co-located with the temperature increase and associated with a preferential perpendicular heating ($T_{p,\perp}>T_{p,\parallel}$), are also observed. 

The ${\bm j}\cdot{\bm E}$ plot indicates that, although this term is not globally connected with temperature increase, it can be locally significant, close to current sheets and regions where kinetic effects are important and intermittent dissipation may be active \citep{osman2011evidence}.

The heat-flux divergence panels (g) support the idea that this term can be locally important, although it does not globally change the proton internal energy. Indeed, in the left-column panel, a strong peak close to current sheets is recovered, while in the right-column panel, $\nabla \cdot {\bm q}_p$ is almost flat. 

Both $-{\bm \Pi}_p : {\bm D}_p$ (g) and $-P_p\theta_p$ (h) report variations across the structures, ranging from positive to negative values. Focusing on the {\it Pi-D} term, it is apparent in the left-column panel, that the positive peak of {\it Pi-D} is displaced from the current structure and it is not associated with the temperature growth. On the other hand, in the right-column panel, two weak negative $-{\bm \Pi}_p : {\bm D}_p$ peaks are found in correspondence of the current structure, while $P_p \theta_p$ slightly increases. 

Figure \ref{fig:1Dcut} helps to realize the complexity of truly understanding how energy is dissipated in a turbulent, weakly-collisional plasma. It is largely established that temperature growth occurs in or around current sheets. The same holds for temperature anisotropy and $\epsilon_p$, although structures are wider. The transfer mechanism based on the {\it Pi-D} term is somehow correlated with hotter plasma regions, although other terms ($P_p\theta_p$, heat-flux) may locally play a role. In our opinion, this should motivate extremely accurate {\it in-situ} measurements to better appreciate the role of these variables. 

\section{Conclusion}
\label{sect:concl}
We have examined the production of internal energy in a collisionless plasma employing an Eulerian HVM scheme and the pressure strain formalism \cite{yang2017energyPOP,yang2017energyPRE,yang2018scale}. 

To our knowledge, a study of this type has not been reported using an Eulerian approach. Relative to previous reports based on PIC models, the HVM approach has the advantage that the proton velocity distribution is well resolved in three dimensions and for several thermal speeds. Therefore we may have some confidence that the subtle interactions between the pressure tensor and the spatial derivative of fluid velocity entailed in the pressure-strain interaction are well-represented in the computed solution. The HVM method is much less ``noisy'' than PIC, especially with regard to particle counting noise associated with finite particle number (see e.g., discussion in \citet{pezzi2017colliding,HaggertyEA17-pop}). This permits HVM to compute accurate structures in velocity space as required for the present study. 

Enabled in this way, we have examined characteristic kinetic properties during turbulence evolution, near the time of maximum mean square current density, when nonlinear effects are strongest. Indeed we have observed at this time the emergence of strong coherent current structures in real space, representing intermittency, and strongly non-Maxwellian features in the proton VDFs, as has been previously reported \cite{GrecoEA12-kinetic,servidio2015kinetic}.

One of the very basic demonstrations of the efficacy of the HVM simulation strategy, and of the physical relevance of the pressure-strain relations, is seen in Fig \ref{fig:timeev}. Here we have shown that the time integration of the sum of the two volume-integrated pressure-strain effects --the pressure-dilatation {\it P-$\theta$} and the {\it Pi-D} terms-- provides a very accurate account of the change of internal energy due to the turbulence cascade. On the other hand, the electromagnetic work on particles, ${\bm j} \cdot {\bm E}$, does not correlate well with global internal energy increases. This would be anticipated based on the analysis leading to Eqs. (\ref{eq:enflowav}-\ref{eq:enmagnav}), as one observes that the charge flux of species interacts with the electric field only to change the species flow energy at the expense of magnetic energy.

The present study has also reported that the values of the pressure-strain energy conversion ($-{\bm \Pi}_p : {\bm D}_{p}$) are broadly distributed and skewed towards positive values associated with production 
of internal energy. Spatial maps furthermore indicate that strong $-{\bm \Pi}_p : {\bm D}_{p}$ values appear
in sheet (or core) like patterns indicating intermittent conversion of flow energy into internal energy. Both of these distributions are typical of quantities related to cascade and pathways to dissipation, such as the``local energy transfer'' (LET) \cite{yang2018scale,sorriso2018statistical,sorriso2019turbulence} or ${\bm j} \cdot {\bm E}$ in various reference frames \cite{Wan:etal:2012,WanEA15}. 

It remains to draw some conclusion regarding the idea of ``dissipation''. A reasonable definition of dissipation may require the conversion of energy into internal energy. Leaving momentarily aside the role of collisions in energy conversion and introduction of irreversibility, the spatial concentration of {\it Pi-D} observed here (and earlier in PIC simulations \citep{yang2017energyPRE,yang2017energyPOP}) appears to be precisely what is required for a plasma physics adaptation of the Kolmogorov Refined Similarity Hypothesis. In such a development, the concentration of dissipation is directly related to scaling of velocity increments, including multi-fractal behavior, etc. In this way, it appears likely that the pressure-strain interaction will enter prominently into more formal and more complete theories of plasma turbulence that remain to be developed. 

The role of collisions in attenuating velocity-space structures and in securing irreversibility remains to be further explored \citep{pezzi2016collisional, pezzi2017solarwind, pezzi2019protonproton}. This need is particularly clear pertaining to the relationship of weak collisions to the strength of pressure-strain interactions, since formal (viscous) closure relating the two is not available in the weakly collisional case. 

Furthermore, within a collisionless system, the internal energy growth is not uniquely related to the increase of the random particle energy, since the internal energy may also increase for the presence of non-Maxwellian features such as accelerated particle beams. The VDF free-energy contained in such structures\citep{lesur2013nonlinear} can be backward released in form of ordered energy, such as micro-instabilities without increasing the plasma entropy\citep{hellinger2017mirror}. Collisions would intrinsically inhibit the possibility to transfer back this from particles to field this energy, since --dissipating these structures-- they incessantly push plasma toward thermal equilibrium \citep{pezzi2016collisional,pezzi2017solarwind}. This further level of complexity should motivate subsequent studies to provide insights on the strength of the physical ingredient that ultimately dissipates energy in a irreversible way\citep{pezzi2019protonproton}.

\begin{acknowledgments}
This research is supported in part by NASA through the MMS Theory and Modeling Program NNX14AC39G and the Parker Solar Probe project under Princeton subcontract SUB0000165, and HSR grant 80SSC18K1648. This paper has received funding from the European Union’s Horizon 2020 research and innovation programme under grant agreement No 776262 (AIDA, www.aida-space.eu). OP and SS are partly supported by the International Space Science Institute (ISSI) in the framework of the International Team 405 entitled ``Current Sheets, Turbulence, Structures and Particle Acceleration in the Heliosphere''. Numerical simulations discussed here have been performed on the Marconi cluster at CINECA (Italy), within the projects IsC53\_RoC-SWT and IsC63\_RoC-SWTB, and on the Newton cluster at the University of Calabria (Italy). 

\end{acknowledgments}

%
\end{document}